\documentclass[12pt]{article}
\usepackage{amssymb}
\newcommand{\be}{\begin{equation}}
\newcommand{\ee}{\end{equation}}

\begin{document}


\begin{titlepage}
 \begin{flushright}
LTH/1135
\end{flushright} 

\vskip 2cm
\begin{center}
{\large \bf  From ${\cal N} =2$ in four dimensions to (0,2) in two dimensions}
\vskip 1.2cm
Radu Tatar 

\vskip 0.4cm

{\it Division of Theoretical Physics, Department of Mathematical Sciences

The University of Liverpool,
Liverpool,~L69 3BX, England, U.K.

rtatar@liverpool.ac.uk}

\vskip 1.5cm

\abstract{We consider ${\cal N}=2$ four dimensional field theories compactified on a two torus in the 
presence of a $U(1)$ magnetic field. We discuss the restrictions leading to theories with (2,2) 
supersymmetry or (0,2) supersymmetry in two dimensions.
The field theories live on  D5 branes wrapped on four cycles of Calabi-Yau 3-folds or 4-folds described as resolved 
ADE singularities or resolved conifold fibered over a two torus.}   

\end{center}
\end{titlepage}


\section{Introduction}
Recently there has been a substantial amount of interest dedicated to study various aspects 
of two dimensional (0,2) theories living on branes. 
The first such brane configurations were introduced about 20 years ago 
\cite{old} as either D1 branes at singularities or brane boxes. 
The recent developments involve using brane tilings and orbifold reductions from D3 probing Calabi-Yau 3-folds to 
D1 probing Calabi-Yau 4-folds \cite{franco}, compactifying F-theory on Calabi-Yau 5-folds 
\cite{sakura} or reducing ${\cal N} = 1$ four dimensional theories on Riemann surfaces 
\cite{benini}. Some important properties like (0,2) trialities have been discovered and their connection to 
${\cal N} = 1$ four dimensional theories opened the gate to exciting developments \cite{gukov}. Other interesting (0,2) theories in two dimensions 
were constructed as $AdS_3$ solutions of M theory and type IIB supergravity 
\cite{gauntlett}.

A different approach was initiated by Kutasov and Lin in 
\cite{kuta1}-\cite{kuta2} with D4 branes stretched between orthogonal NS branes. The field theory on 
D4 branes is    ${\cal N} = 1$ 
in four dimensions which is further compactified on a two torus to get a two dimensional (2,2) theory. The 
supersymmetry can be broken by either turning on a  D-term or a magnetic flux on the two torus. 
When both are considered and the  D-term is equal to the magnetic flux,  the supersymmetry is partially 
preserved and one gets a two dimensional (0,2) theories. The value of the  D term and the magnetic field  can be 
read from the various rotations of the D4 branes. 

The T-dual picture of the Kutasov-Lin results was considered in 
\cite{tatar} based on the T-duality between brane configurations 
with D4 branes stretched between orthogonal NS branes and D5 branes wrapped on 2-cycles of resolved conifold configurations \cite{rt}.  

A non-zero D-term corresponds to rigid $\mathbb{P}^1$ cycles \cite{aga1}. To obtain two dimensional theories the D5 branes 
are wrapped on four cycles which are $\mathbb{P}^1$ fibres over the two torus. The equality between the D-term and the magnetic field appears 
when imposing the covariant spinor condition on the wrapped D5 branes.

In this work we go one step further and consider a larger class of two dimensional (0,2) theories arising from 2-torus  compactification  of ${\cal N} =2$ 
four dimensional theories with non-zero magnetic field and non zero D term. The four dimensional field theories live on D5 branes wrapped on resolution cycles of ALE spaces.  If the D-term and the magnetic field are equal, we show that the two dimensional SUSY
is broken from (4,4) to (2,2). Adding a superpotential as a polynomial in the chiral adjoint field breaks 
${\cal N} = 2$ to ${\cal N} = 1$ in four dimensions and we see that the two dimensional supersymmetry is broken from (2,2) to (0,2). 

There are various aspects covered in this work:

1) in section 2 we consider the ALE spaces and their resolutions fibered over a two torus. For 
$A_1$ singularity and its resolutions, one interesting observation is that the patches covering the resolution 
$\mathbb{P}^1$ are square root fibre bundles and this restricts the model to tensor products of even degree. In the present work we only consider models with fibres of degree zero when the fibre group becomes the zero Picard group which is a dual torus. This has four  2-torsion  points and each 
2-torsion point gives rise to a different (0,2) theory.

2) in section 3 we consider the existence of covariant spinors on the  wrapped D5 branes on 
four cycles of Calabi-Yau 3-folds and 4-folds. The supersymmetry preservation implies a geometric
 equality between the magnetic flux and D-term, also involving the NS flux through the  $\mathbb{P}^1$ cycle and 
the area of the two torus. 

3) In section 4 we discuss how the multiplets of the four dimensional  ${\cal N} =2$ theory reduce to  
multiplets of a (2,2) supersymmetric theory in two dimensions when compactified on a two torus. 
In the presence of non zero D-terms and magnetic flux the supersymmetry is partially broken to (2,2) in two dimensions. 
The ${\cal N} =2$
supersymmetry in four dimensions is broken to  ${\cal N} =1$ by adding a superpotential for the chiral adjoint field. 
When reducing to two dimensions on a two torus with magnetic flux, the superpotential breaks (2,2) supersymmetry to 
(0,2) supersymmetry. 

4) In section 5 we consider the deformations of brane configurations which map a (4,4) two dimensional theory into a (0,2) two dimensional theory. The (4,4) two dimensional theory lives on  D4 branes suspended between parallel NS branes. By rotating the 
D4 branes one reaches the (2,2) supersymmetric theory and a further rotation of the NS branes provide the (0,2) theory. 

This work is a first step towards building a large set of (0,2) two dimensional theories. There are many result in \cite{rt} which 
can be reconsidered after compactification on a two torus.

\section{Calabi-Yau manifolds as fibrations over $T^2$}
We start by describing the Calabi-Yau 3-folds and 4-folds as fibration of singular spaces and their resolutions over a two torus.

\subsection{Calabi-Yau threefolds}
In this subsection we consider the case of Calabi-Yau 3-folds as ALE spaces and their resolutions fibered over a 
two torus. One important aspect is the appearance of square root line bundles which require a careful treatment of 2-torsion line bundles.

Consider the $A_1$ singularity in $\mathbb{C}^3$ i.e. the space $\mathbb{C}^2/Z_2$. This can be 
embedded in $\mathbb{C}^3(x_0,x_1,x_2)$ as a hypersurface 
\be
\label{a}
A: x_0~x_1 - x_2^2 = 0
\ee
 which has a singularity at the origin. To smooth it out we blow-up $\mathbb{C}^3$ at the origin by replacing (0,0,0) with an exceptional divisor  
$\mathbb{P}^2$. When we follow a path in the hypersurface $A$ towards the origin, we land  on the exceptional $\mathbb{P}^2$
in the blow-up which provides a set $[X_0,X_1,X_2]$ on $\mathbb{P}^2$ related by the quartic $X_0~X_1 - X_2^2 = 0$, which is isomorphic to  
a $\mathbb{P}^1$ resolution cycle. 

The resolution $\mathbb{P}^1$  cycle can be wrapped by D5 branes to provide an  ${\cal N}= 2$ SUSY theory in four dimensions. The normal bundle to  $\mathbb{P}^1$ in the resolved space is $O(-2) \oplus O(0)$. The $O(0)$ coordinate is denoted as $X$ and the 
$O(-2)$ part can be understood as following: the $\mathbb{P}^1$  is covered by two affine pieces, one with $X_1 \ne 0$ parametrised by $\xi = X_0/X_2$ and one with $X_0 \ne 0$ parametrised by $\eta = X_1/X_2$. The gluing between the two pieces is given by 
\be
\label{affinec}
\eta = \xi^{-1}, X_0 = X_1 \xi^2
\ee

When $\mathbb{C}^2/Z_2$ is fibered over a two torus the complex coordinates
of $\mathbb{C}^3(x_0,x_1,x_2)$ are promoted to being sections of line bundles $L_0,L_1,L_2$ over the torus and the 
local Calabi-Yau threefold is given by relation (\ref{a}) in the four dimensional complex variety 
$L_0 \oplus L_1 \oplus L_2 \rightarrow T^2$.  The line bundles obey
$L_0 \otimes L_1 = L_2 \otimes L_2$. 

We now consider  the resolution of the $\mathbb{C}^2/Z_2$ singularity as the fibre in a Calabi Yau 3-fold. 
The coordinates  
$x_0, x_1, x_2$ are replaced by $X_0, X_1, X_2$ and we use the same notation for their interpretations as line bundles: $L_0, L_1, L_2$. The affine coordinates of the two coordinates patches of the $\mathbb{P}^1$ fibres  $\eta$ and $\xi$ given in 
(\ref{affinec})  are sections of the line bundle $L$ satisfying the condition 
$L_0 \otimes L_1^{-1} = L \otimes L$ for $\xi$ and  
$L_1 \otimes L_0^{-1} = L \otimes L$ for $\eta$. In general a line bundle 
has a non-zero degree which is the number of zeroes minus the number of poles 
in any holomorphic section. For simplicity, in the current work we limit to the case of zero degree. If we consider 
the complex surface $S$ obtained by 
fibering  $\mathbb{P}^1$ over $T^2$, for zero degree  line bundle the volume of $S$ is given by the product of the volume of the 
 $\mathbb{P}^1$ fibre and the area of the $T^2$ base. 

 A formula like $L_0 \otimes L_1^{-1} = L \otimes L$ implies that the line bundle $L$ is the square root of the tensor product of line bundles $L_0 \otimes L_1^{-1}$. 
If it exists, the square root line bundle is not unique in general and 
two square root line bundles differ by a 2-torsion line bundle. 

We briefly remind what the 2-torsion point of a two torus are. The 2-torus is the quotient 
$T^2 =\mathbb{C} /\Lambda$ obtained by dividing the complex plane by a lattice $\Lambda = \mathbb{Z}^2$.  The torus admits the 
involution $x \rightarrow - x$ and for every $y$ also admits the involution $x \rightarrow 2 y - x$, which fixes $y$.
The question is for which $y$ the combined result of the involutions takes $x$ into $x$. This would mean that 
$2 y - x  = - x$ or $ 2 y = 0$. This is the definition of 2-torsion points on the torus and there are 4 such points for a two torus.    

The Picard group is the group of holomorphic line bundles. In case of line bundles of zero degree, one deals with the zero 
Picard group $Pic^0(T^2)$ which is a dual torus and has four 2-torsion points. The 2-torsion condition 
is written as $\tilde{L} \otimes \tilde{L} = E$ where is $E$ is a trivial bundle. So once the  product $L_0 \otimes L_1^{-1}$
is defined as $L \otimes L$, there are 4 different values for $L \otimes \tilde{L}$ which can't be distinguish when considering  $L \otimes L$.

The $A_1$ discussion can be generalised to any A-D-E singularity which can be blown up to a smooth space where 
the singular point is replaced by a collection of rational curves $\mathbb{P}^1_i$. For an $A, D, E$ group
of rank $n$, there is a collection of $n$  $\mathbb{P}^1$ cycles,  each having an $O(-2)$ 
fibre bundle. The total space of the normal bundle over the $i$-th  $\mathbb{P}^1_i$ is  
\be
\label{affinecm}
\eta_i = \xi_i^{-1}, X_{0i} = X_{1i} \xi_{i}^2,~~i=1,\cdots,n
\ee    
The D5 branes are wrapped on $n$ complex surfaces $S_i$ obtained by fibering the $n$ various  $\mathbb{P}^1_i$ over 
$T^2$.  In case we have line bundles of degree 0, the volumes of each $S_i$ 
is the product of the volume of the  $\mathbb{P}^1_i$ fibre and the volume of the
$T^2$ base. 

$X_{0i}$ corresponds to a line bundle $L_{0i}$, $X_{1i}$ to a line bundle $L_{1i}$ and $\xi_i$ to a line bundle 
$L_{i}$. The consistency condition requires that for each $\mathbb{P}^1_i$ 
\be
L_{0i} \otimes L_{1i}^{-1} = L_i \otimes L_i.
\ee
Each $L_i$ comes with its own 2-torsion point in the dual torus.  

\subsection{Calabi-Yau fourfolds}

The Calabi-Yau fourfolds as resolved/deformed conifold singularities fibered over $T^2$ were considered in \cite{int1,tatar}. 

The Calabi-Yau fourfolds are obtained by fibering conifold type geometries  $x_0 x _1 = x_2 x_3$ over two tori. The $x_i, i =0,1,2,3$ become line bundles $L_i, i =0,1,2,3$ satisfying the condition
\be
L_0 \otimes L_1 = L_2 \otimes L_3 
\ee

The conifold singularity is resolved by replacing the origin with  a  $\mathbb{P}^1_i$ with homogeneous coordinates $u, v$ such that $x_0 u = x_2 v$ and the affine coordinates of the two
patches of the  $\mathbb{P}^1_i$ fibre are $z = u/v, w = v/u$. They are sections of 
$L_0 \otimes L_2^{-1}$ and $L_2 \otimes L_0^{-1}$ respectively. 

When deforming the Calabi-Yau threefolds into Calabi-Yau fourfolds, we need to also consider the change in the equation 
(2). The simplest deformation of the fibre is
\be
\label{fourf}
X_0 = X_1 \xi^2 + X \xi
\ee
where $X$ is the coordinate of the $O(0)$ part of the normal bundle inside the Calabi-Yau threefold. We see that the consistency condition requires
\be
L_0 \otimes L_1^{-1} = L \otimes L
\ee
and 
\be
L_0 = L_{X} \otimes L
\ee
We see that line bundle over $X$ can be written in terms of $L_0, L_1, L$ so depends on the choice of the 2-torsion point inherited form the Calabi-Yau threefold. 
It would be interesting to see the dependence of the two dimensional field theories on the choice of the 2-torsion point. 

\section{Field theory on wrapped D5 branes}

\subsection{D5 branes wrapped on 2-cycles of Calabi-Yau 2-folds}

We start with D5 branes wrapped on 2-cycles of SU(2) holonomy manifolds. The 
sizes of the 2-cycles are non-vanishing due to either having a non-zero NS field or a real Kahler modulus. In reality there 
exists a three dimensional space of deformations of the ALE metric for each 
of the  $\mathbb{P}^1$ cycles:

-  a complex parameter $\alpha$ corresponding to integral of the holomorphic 2-form over the  $\mathbb{P}^1$.

- a real parameter $j$ corresponding to the integral of Kahler form on the  $\mathbb{P}^1$. 

Together with the integral of the NS field $B_{NS}$ on $\mathbb{P}^1$,
$b_{NS} = \int_ {\mathbb{P}^1} B_{NS}$,  we get  a 5 parameter family of deformations and the stringy volume is 
$(j^2 + b_{NS}^2 + |\alpha|^2)^{1/2}$. In terms of the field theory parameters, 
$\alpha$ corresponds to a field theory F-term and $r$ to a field theory D-term. In the current work we limit to the discussion of SUSY breaking when turning on 
D-terms so we set $\alpha = 0$ and the volumes becomes 
$(j^2 + b_{NS}^2)^{1/2}$.

The two important limits of $(j^2 + b_{NS}^2)^{1/2}$ are $j=0$ which corresponds to a fractional D3 brane and $b_{NS} = 0$ which corresponds to D5 branes wrapping rigid 2-cycles. The two solutions can be interpolated by uplifting to M theory and performing boosts \cite{maldacena-martelli}. The boost along $t$ and $x^{11}$ direction is 
\be
t \rightarrow \mbox{cosh} \beta~t - \mbox{sin} \beta~x_{11}, x_{11} \rightarrow -   \mbox{sinh} \beta~t~\mbox{cos} \beta~z_{11}
\ee
The boost parameter $\beta$, the NS field $B_NS$, the Kahler volume $J$ and the dilaton are related as
\be
\label{rel1}
B_{NS} =  \mbox{sinh} \beta e^{- 2 \Phi} J.
\ee
The ${\cal N}=1$ interpolating solution between rigid branes and fractional branes  of \cite{maldacena-martelli} has been generalised to ${\cal N}=2$ models 
in \cite{minasianetal}. For 
a constant dilaton $\Phi = \Phi_0$, (\ref{rel1})  can be integrated over  $\mathbb{P}^1$ and implies 
\be
b =  \mbox{sinh} \beta  e^{- 2 \Phi_0} j. 
\ee
The calibration conditions do not change if we turn on magnetic flux or NS flux. For D5 branes wrapping a  2-cycle inside an 
$SU(2)$ holonomy manifold with no magnetic or NS flux, the solution corresponds to the usual calibration condition of a 2-cycle in a K3 manifold 
given by the condition
\be
\label{calcond}
(J, Re(\Omega), Im (\Omega) = (\mbox{cos} \theta,\mbox{sin} \theta~\mbox{cos} \phi,  \mbox{sin} \theta~\mbox{sin} \phi) 
vol_{\mathbb{P}^1}
\ee
where $\theta$ and $\phi$ are constant angles along the two-cycle. We can add to $J$ two types of antisymmetric tensors in two dimensions, one is the magnetic flux $M$ and the other is the $NS$ field. As discussed in \cite{minasian}, the antisymmetric tensors do not change the calibrations conditions and 
the supersymmetry is preserved. The only difference is that instead of $J = \mbox{cos} \theta~vol_{\mathbb{P}^1}$, we would have 
$J + i B_{NS} = e^{i \theta}  ~vol_{\mathbb{P}^1}$ and $J~\mbox{tan} \theta = B_{NS}$ such that the coupling constant of the field theory on the wrapped D5 branes is 
\be
j ~\mbox{cos} \theta + b ~\mbox{sin} \theta = \sqrt{j^2+b^2}.
\ee
As $\theta$ is related to the boosting parameter $\beta$, we also see that for D5 branes wrapping two $\mathbb{P}^1$ cycles with values $b_1, b_2$ for
$\int_{\mathbb{P}^1_i} B_{NS}$ and  $j_1, j_2$ for $\int_{\mathbb{P}^1} J$, the two cycles should be calibrated such that 
\be
\frac{b_1}{j_1} = \frac{b_1}{j_1},
\ee
otherwise the supersymmetry is fully broken.  

\subsection{D5 branes wrapped on 4-cycles of Calabi-Yau 3-folds}

We now want to consider the case when the supersymmetry is preserved in the presence of both NS flux $B_{NS}$ and magnetic flux
$M$. To do this we consider a four cycle 
$S$ as a nontrivial $\mathbb{P}^1$ fibration over $T^2$. The condition of preserving supersymmetry when D5 branes wrap four-cycles is 
\be
(J_{\mathbb{P}^1} + i B_{NS}) (A_{T^2} + i M) = e^{i \theta} \frac{\sqrt{|g+M|}}{\sqrt{|g|}} vol_{S}
\ee
where $vol_{S}$ is the volume of the four cycle. 

This relation can be split into a real part
\be
\label{real}
J_{\mathbb{P}^1}~A_{T^2} - B_{NS}~M = \mbox{cos}~\theta  \frac{\sqrt{|g+M|}}{\sqrt{|g|}} vol_{S}
\ee
and an imaginary part 
\be
\label{imaginary}
J_{\mathbb{P}^1}~M + B_{NS}~A_{T^2} = \mbox{sin}~\theta  \frac{\sqrt{|g+M|}}{\sqrt{|g|}} vol_{S}
\ee
When the magnetic flux and the NS flux $B_{NS}$ are set to zero, the  equations 
(\ref{real})-(\ref{imaginary}) are satisfied for $\theta=0$ which reduces to the original
condition that the four cycle is holomorphic 
\be
J_{\mathbb{P}^1}~A_{T^2} = vol_{S}. 
\ee
We see that the equation (\ref{real}) is also satisfied when 
$J_{\mathbb{P}^1}~A_{T^2} = B_{NS}~M$ for $\theta = \frac{\pi}{2}$. This is the geometric version of the equality between the D-term and the magnetic flux which first appeared in \cite{kuta1}.

Our result is that the four cycle $S$ is holomorphic 
when $J_{\mathbb{P}^1}~A_{T^2} = B_{NS}~M$. This is the geometrical relation between 
the rigidity parameter $J_{\mathbb{P}^1}$ for the $\mathbb{P}^1$ cycle and the magnetic flux $M$ 
through $T^2$ and is the equivalent of the equality between the 
D term and the magnetic flux in field theory, considered in the next section.   
    
\subsection{D5 branes wrapped on 4-cycles of Calabi-Yau four-folds}

As we aim to describe supersymmetric theories with (0,2) supersymmetry in 2 
dimensions, we consider wrapping D5 branes on 4-cycles of Calabi-Yau four-folds. We consider the Calabi-Yau fourfolds as 
resolved conifold geometries fibered over a two torus.

Fortunately the result of the previous subsections and the ones of \cite{tatar} allow us to directly build 
these geometries. We turn on a magnetic flux $M$ and make the $\mathbb{P}^1$ cycle 
rigid such that  $J_{\mathbb{P}^1}~A_{T^2} = B_{NS}~M$ which ensures (2,2) supersymmetry in two dimensions. The second step is to change the 
$\mathbb{P}^1$ normal bundle from 
$O(0) \oplus O(-2)$ to $O(-1)  \oplus O(-1)$ without changing the 
  $J_{\mathbb{P}^1}~A_{T^2} = B_{NS}~M$ relation which ensures the preservation of 
(0,2) supersymmetry. 

The set-up can be made more complicated if 
the processes of modifying the normal bundle to $\mathbb{P}^1$, turning on the 
magnetic flux and making the cycle rigid are all done at the same time. We plan to consider this more general consideration in a future publication.

\section{Field Theory: from ${\cal N} = 2, d=4$ to $(0,2), d=2$}

\subsection{${\cal N} = 2, d=4$ Theory}

\subsubsection{Theory without Flavours}

Consider an ${\cal N} = 2, d=4$ theory with a gauge group $SU(N_c)$ and no flavours. An  ${\cal N} = 2$ vector 
multiplet consists of an   ${\cal N} = 1$ vector multiplet 
$(\lambda, A_{\mu})$ and an   ${\cal N} = 1$ chiral multiplet 
$(\phi, \psi)$ in the adjoint representation of the  group $SU(N_C)$. We denote the ${\cal N} = 1$ chiral multiplet by $\Phi$. The two 
supersymmetry transformations are:

- the first SUSY transformation acts inside the  ${\cal N} = 1$ vector or chiral multiplets and relates $\lambda$ to $A_{\mu}$ and  
$\phi$ to $\psi$ respectively. 

- the second SUSY transformation is obtained by rotating the fermions 
$\lambda$ and $\psi$ into each other 
\be
\label{rot}
\lambda \rightarrow i~\psi;~~\psi~\rightarrow~- i ~\lambda
\ee

\subsubsection{Theory with Flavours}

We now add flavour fields as ${\cal N} = 2$ hypermultiplets. An 
${\cal N} = 2$ hypermultiplet consists of an  ${\cal N} = 1$ chiral multiplet
$(Q, \phi_{Q})$ and an  ${\cal N} = 1$ antichiral multiplet 
$(\tilde{Q}^{\dagger}, \phi_{\tilde{Q}}^{\dagger})$. 
$Q$ and $\tilde{Q}^{\dagger}$ are in  the same representation of the gauge 
group which implies that  $Q$ and $\tilde{Q}$ are in conjugate representation 
of the gauge group.  For hypermultiplets the two 
supersymmetry transformations act as:

- the first SUSY transformation acts inside the two ${\cal N} = 1$ chiral 
multiplets and connects $Q$ with $ \phi_{Q}$ and  $(\tilde{Q}^{\dagger}$ and
$\phi_{\tilde{Q}}^{\dagger})$ respectively. 

- the second SUSY transformation is obtained by rotating $Q$ and 
 $\tilde{Q}^{\dagger}$ into each other. 

For $SU(N_c)$ gauge group with $N_f$ flavours in the fundamental representation,
$Q$ and $\tilde{Q}^{\dagger}$ are in the fundamental representation which implies 
that  $\tilde{Q}$ is in the antifundamental representation. The coupling between  the matter fields and the ${\cal N} = 2$ vector multiplet is  written in the  ${\cal N} = 1$ superspace as 
$\int d^2~\theta~ Q~\Phi~\tilde{Q}$.

\subsection{From ${\cal N} = 2, d=4$ to $(2,2), d=2$}  
\subsubsection{Equal magnetic flux and D term lead to   $(2,2), d=2$}

 We now want to discuss the breaking of supersymmetry to (2,2) in two dimensions. To do this we apply a similar procedure 
to the one in \cite{kuta1} but when the starting point is an  ${\cal N} = 2, d=4$ theory instead of 
 ${\cal N} = 1, d=4$ theory. We consider that the flavours are charged with respect to an external $U(1)$ group. The fields $Q$ and $\tilde{Q}^{\dagger}$ have the same charge under the $U(1)$ group which implies that the fields 
 $Q$ and $\tilde{Q}$ have opposite charges. The current supermultiplets of the 
  ${\cal N} = 2, d=4$ theory are coupled to  an external  ${\cal N} = 2, d=4$
vector multiplet with the fermion content $\lambda$ and $\psi$.

Consider that the four dimensional theory lives in the $(x^0,x^1,x^2,x^3)$ space and the directions $(x^1,x^2)$ are 
compactified on a two torus. We chose the expectation value of the magnetic field through the torus to be
$F_{12} = M$ which breaks both supersymmetries
\be
\delta_{\xi} \lambda =  F_{\mu \nu} \sigma^{\mu \nu} \xi,~~
\delta_{\xi'} \psi = - i F_{\mu \nu} \sigma^{\mu \nu} \xi'
\ee

The second equation arises from the map 
$\lambda \rightarrow i \psi$ and $\psi \rightarrow - i \lambda$  when  interchanging the two SUSY transformations 
generated by $\xi$ and $\xi'$.  
  
To preserve supersymmetry, we turn on a non-zero D field for the external $U(1)$ group and we encounter the super Bogomolnyi limit of 
the model treated in \cite{achucaro}. 
They considered ${\cal N} = 1, d=4$ supersymmetric QED theory with three chiral superfields $\Phi_0,\Phi_+, \Phi_-$ with charges 0, 1, -1. In our case the field $Q$ has positive charge and takes the place of $\Phi_+$,  $\tilde{Q}$ has negative charge and 
takes the place of $\Phi_-$ whereas the ${\cal N} = 1$ chiral multiplet component of the   ${\cal N} = 2$ vector multiplet is uncharged and replaces  $\Phi_0$. 

The transformations of $\lambda$ and $\psi_0$ components of the  ${\cal N} = 1$ gauge multiplet and the  ${\cal N} = 1$ 
neutral chiral multiplet are \cite{achucaro}
\be
\delta_{\xi} \lambda =  (F_{\mu \nu} \sigma^{\mu \nu} + i D) \xi
\ee
and 
\be
\delta_{\xi'} \psi_0 = (- i F_{\mu \nu} \sigma^{\mu \nu} -  D) \xi' 
\ee
$D$ field is related to the charged scalar fields due to the presence of the terms $D^2 + D \phi^{\dagger} T \phi$ in the Lagrangian which
 gives $D = |\phi_+|^2 -  |\phi_-|^2$.

Consider now the resulting two dimensional  spacetime of the form $R^{1,1} \times T^2$ with $R^{1,1}$ spanned by $(x^0,x^3)$ and $T^2$ described by the coordinates $(x^1,~x^2)$. The  
magnetic flux has zero components only through the 2-torus, $F_{12} = B$. The 4-dimensional spinors $\xi$ and $\xi'$ are doublets 
$(\xi_-,\xi_+), (\xi'_-, \xi'_+)$ of right and left $R^{1,1}$ spinors. The 2-dimensional supersymmetry is (4,4) and 
the possible partially broken
supersymmetry could be (4,0), (0,4) or (2,2) supersymmetry in two dimensions. To see which one is actually obtained,  
we first consider the choice $D = B$ which 
implies that  the theory preserves (0,2) SUSY coming from $\xi$ and (2,0) SUSY coming from $\xi'$, they combine to provide a (2,2) SUSY.  
For the case $D = - B$ the theory preserves (2,0) SUSY coming from $\xi$ and (0,2) SUSY coming from $\xi'$ which combine to 
provide a (2,2) SUSY. We conclude that for $D=\pm B$, the two supersymmetries are partially broken and we remain with (2,2) supersymmetry in two dimensions. It is not clear how to obtain (0,4) or (4,0) models ( see \cite{gukov1,tong} for considerations of such examples).

\subsubsection{Fields in  the  $(2,2), d=2$ theory}
We now discuss the surviving fields and superpotential in the $(2,2), d=2$ theory.  The starting point is a (4,4) two dimensional 
theory which is obtained by 
reducing a four dimensional ${\cal N} = 2$ theory on a torus. The  ${\cal N} = 2$ theory has  ${\cal N} = 2$ vector multiplets  and ${\cal N} = 2$ hypermultiplets .  When reduced to the  (4,4) two dimensional theory,
the four dimensional vector multiplet decomposes into a $(2,2), d=2$ chiral multiplet denoted by $\Phi$ and a twisted chiral mutiplet $\lambda$. 
The four dimensional chiral multiplet decomposes into two (2,2) chiral multiplets $Q$ and $\tilde{Q}$ in conjugate representations of the gauge group. 

There is also  a superpotential $\tilde{Q} \Phi Q$ integrated over the (2,2) superspace $\theta_{1,+} \theta_{1,-}$. We can write the (2,2) multiplets in 
terms of  (0,2) components
\be
Q = Q^{(0,2)} + \sqrt{2} \theta^- \Lambda^{(0,2)}_{Q} - i \theta^{-} \bar{\theta}^{-} (D_0 - D_3) Q^{(0,2)},
\ee
\be
\tilde{Q} = \tilde{Q}^{(0,2)} + \sqrt{2} \theta^- \tilde{\Lambda}^{(0,2)}_{\tilde{Q}} - i \theta^{-} \bar{\theta}^{-} (D_0 - D_3) \tilde{Q}^{(0,2)},
\ee
\be
\Phi = \Phi^{(0,2)} + \sqrt{2} \theta^- \Lambda^{0,2}_{\Phi} - i \theta^{+} \bar{\theta}^{+} (D_0 + D_3) \Phi^{(0,2)},
\ee
where $Q^{(0,2)}, \tilde{Q}^{(0,2)}$ and $\Phi^{(0,2)}$ are (0,2) chiral superfields and $\Lambda^{(0,2)}_{Q}, \tilde{\Lambda}^{(0,2)}_{\tilde{Q}}$
and $\Lambda^{0,2}_{\Phi,+}$ are the corresponding (0,2) Fermi superfields. The (0,2) superfields have themselves expansions as 
\be
Q^{(0,2)} = q + \sqrt{2} \theta^+ \psi_{+} - i \theta^{+} \bar{\theta}^{+} (D_0 + D_3) q,
\ee
\be
\tilde{Q}^{(0,2)} = \tilde{q} + \sqrt{2} \theta^+ \tilde{\psi}_{+} - i \theta^{+} \bar{\theta}^{+} (D_0 + D_3) \tilde{q},
\ee
\be
\Phi^{(0,2)} = \phi + \sqrt{2} \theta^+ \psi_{\phi,+} - i \theta^{+} \bar{\theta}^{+} (D_0 + D_3) \phi,
\ee
and similar ones for the Fermi superfields. 

We now discuss what happens when the $(4,4), d=2$ SUSY is broken to $(2,2), d=2$ due to the combined effect of turning on a 
magnetic field and a D term.  To do this we need to consider the effect on the   ${\cal N} = 2$ hypermultiplet. This contains an   ${\cal N} = 1$
chiral multiplet with scalar component $Q$ and  an   ${\cal N} = 1$ antichiral multiplet with scalar component $\tilde{Q}^{\dagger}$. Both $Q$ and
$\tilde{Q}^{\dagger}$ transform in the same representation of the gauge group. If we turn on an extra $U(1)$ gauge group,  $Q$ and
$\tilde{Q}^{\dagger}$ are required to have the same charge $e$ which means that $\tilde{Q}$ has charge $-e$. The vector multiplet for the original gauge 
group is not charged under the extra $U(1)$ so the superpotential $\tilde{Q} \Phi Q$ has zero charge, as it should.   

 We now identify the massless fields present in the two dimensional (2,2) theory after the first step of SUSY breaking,
when the contributions of the magnetic field and the D-term are taken into account. As in \cite{kuta1}, we consider a 
free complex scalar charged under a $U(1)$ gauge field which is represented by a background gauge field $A_2 = B~x_1$. 
The Klein-Gordon equation for $\phi$ corresponds to the Landau problem for a particle in magnetic field
with a mass spectrum
\be
m_{n}^2 = (2~n+1) |e~B|.
\ee
which can be made zero by turning on the D component of the vector multiplet, leading to the mass spectrum
\be
m_{n}^2 = (2~n+1) |e~B| - e~D
\ee
For $B > 0$ we see that the fields with positive charge $e>0$ give rise to massless two dimensional scalars whereas the 
ones with negative charge $e<0$ do not. Nevertheless, there are also spin 1/2 field with a spectrum
\cite{kuta1}
\be
m_+^2 = (2 n+ 1) |e B| - e~B, m_-^2 = (2 n+ 1) |e B| + e~B.
\ee
We see that the right moving fermions can be massless for positive $e$ and for $B=D$ they can be combined with the corresponding 
massless scalars to give rise to (0,2) chiral multiplets. Therefore the 4 dimensional fields $Q$ reduce to two dimensional 
(0,2) chiral multiplets denoted by $\Phi_{Q}^{e>0}$ with an expansion
\be
\Phi_{Q}^{e>0} = \phi_{Q,+}^{e>0} + \sqrt{2} \theta^{+} \psi_{Q,+}^{e>0} - i \theta^+ \bar{\theta}^+ (D_0 + D_3) \phi_{Q}^{e>0}.
\ee
where $\phi_{Q}^{e>0}$ is the massless complex scalar field and $\psi_{Q,+}^{e>0}$ is the corresponding massless complex 
right-moving fermion. 

Due to the absence of a massless two dimensional scalar from the spectrum for negative magnetic charge, a four 
dimensional fields $\tilde{Q}$ with negative magnetic charge reduces to a two dimensional (0,2) Fermi superfields with the following expansion:
\be
\Lambda_{\tilde{Q}}^{e>0} = \psi_{\tilde{Q}}^{e>0} - \sqrt{2} \theta^+ F^{e>0} 
- i~\theta^+ \bar{\theta}^+ (D_0 + D_3)  \psi_{\tilde{Q}}^{e>0} 
- \sqrt{2}   \bar{\theta}^+ E^{e>0}
\ee
where $E$ is a chiral superfield which is combination of other chiral superfields in the theory. 
The fields $\Phi_{Q}^{e>0}$ and $\Lambda_{\tilde{Q}}^{e>0}$ combine into a (2,2) chiral multiplet. 

The chiral multiplet inside the   ${\cal N} = 2, d=4$ vector multiplet is not charged under the magnetic field and 
will be a full (2,2) two dimensional chiral multiplet with the following $\theta^-$ expansion 
\be
\Phi^{2,2} = \Phi_{\Phi} + \sqrt{2} \theta^{-} \Lambda_{\Phi} - i~\theta^+ \bar{\theta}^+ (D_0 - D_3)  \Phi_{\Phi}
\ee
where $\Phi_{\Phi}$ is a (0,2) chiral superfield and  $\Lambda_{\Phi}$ is a (0,2) Fermi superfield.  

The ${\cal N} = 2, d=4$ theory with $N_f$ flavours has a $SU(N_f)$ flavour symmetry and a $U(1)_R \times SU(2)_R$ 
R-symmetry. The ${\cal N} = 1, Q$ components of the ${\cal N} = 2$ hypermultiplet transform in the fundamental 
representation of $SU(N_f)$ and the  ${\cal N} = 1, \tilde{Q}$ components in the antifundamental representation of $SU(N_f)$.
We chose the supersymmetry breaking magnetic field to represent a $U(1)$ group inside $SU(N)_f$. The $U(1)$ charges for the
 ${\cal N} = 1$ components of the i-th  ${\cal N} = 2$ hypermultiplet  
are related by $e_i = - \tilde{e}_i$ where $e_i$ is the 
charge of $Q_i$ and $\tilde{e}_i$ the charge of $\tilde{Q}_i$. The global symmetry would then be an $SU(N_f)$ acting on the 
$Q$ fields times an $SU(N_f)$ acting on the  $\tilde{Q}$ fields. 
   
But this is not the fully story. As discussed in \cite{kuta1}, the anomaly freedom constraints for a global $U(1)$ orthogonal to 
the gauge group require an extra condition on $e_i$ and $\tilde{e}_i$:
\be
\sum_i e_i = \sum_i \tilde{e}_i = 0
\ee
and we use the same choice as in \cite{kuta1} to take $N_f/2$ of the $e_i$ and $N_f/2$ of the $\tilde{e}_i$ to be $+1$ and
the rest to be $-1$. The global symmetry is in general broken to $SU(N_f/2)^4 \times U(1)$ but there is a  superpotential
inherited from the  ${\cal N} = 2$ theory
\be
\int d \theta  (Q^{e=1} \Phi \tilde{Q}^{e=-1} + Q^{e=-1} \Phi \tilde{Q}^{e=1})
\ee
which breaks the global symmetry from  $SU(N_f/2)^4$ to  $SU(N_f/2)_1 \times SU(N_f/2)_2$. 
The fields $Q^{e=1}$ and $ \tilde{Q}^{e=-1}$ belong to the fundamental and anti-fundamental representations of $SU(N_f/2)_1$ whereas 
 $Q^{e=-1}$ and $ \tilde{Q}^{e=1}$ belong to the fundamental and anti-fundamental representations of $SU(N_f/2)_2$. The field $\Phi$ is not 
charged under   $SU(N_f/2)_1 \times SU(N_f/2)_2$. 

The $\theta^-$ expansions for the reduction of the field $Q^{e=+1}$ and $Q^{e=1}$ (or $\tilde{Q}^{e=+1}$ and $\tilde{Q}^{e=1}$) are  
\be
Q^{2,2} = \Phi_{Q},~~ \tilde{Q}^{2,2} = \sqrt{2} \theta^{-} \Lambda_{\tilde{Q}},
\ee
due to the absence of the (0,2) Fermi superfield and (0,2) chiral superfield respectively.  

We can now see what is the reduction of the four dimensional superpotential $\int d^2~\theta~Q~\Phi~\tilde{Q}$. In the 
(2,2) two dimensional theory notations this would be
\be
\int~d~\theta^+ d~\theta^-  \Phi_{Q} (\Phi_{\Phi} + \sqrt{2} \theta^{-} \Lambda_{\Phi})  \sqrt{2} 
\theta^{-} \Lambda_{\tilde{Q}}
\ee
The integration over $\theta_-$ provide the superpotential
\be
\label{supot1}
\int~d~\theta^+  \sqrt{2} \Phi_{Q} \Phi_{\Phi}  \Lambda_{\tilde{Q}}+ 
\ee

The original  ${\cal N} = 2, d=4$ theory has both 
 $Q$ and  $\tilde{Q}^{\dagger}$ in the same representation of the gauge group. They also have the same charge +1 under the
extra global $U(1)$ group.  At the same time,  $Q^{\dagger}$ and  $\tilde{Q}$ also have the same charge -1 under extra global $U(1)$
group. The previous argument implies that, after the reduction to 2 dimensions, the field   $Q^{\dagger}$ reduces to a 
(0,2) Fermi field  $\Lambda_{Q}$ whereas  $\tilde{Q}^{\dagger}$ reduces to a (0,2) chiral superfield $\Phi_{\tilde{Q}}$.
The fields 
$\Phi_{Q}$ and  $\Lambda_{Q}$ together form a (2,2) multiplet
\be
\label{exp1}
\Phi_{Q}^{(2,2)}= \Phi_{Q} + \sqrt{2} \Lambda_{Q} - i~\theta^{-} 
\bar{\theta}^{-} (D_0 - D_3)  \Phi_{Q},
\ee
and  $\Phi_{\tilde{Q}}$ and $\Lambda_{\tilde{Q}}$ also form a (2,2) multiplet
\be
\label{exp2}
\Phi_{\tilde{Q}}^{(2,2)}= \Phi_{\tilde{Q}} + \sqrt{2} \Lambda_{\tilde{Q}} - i~\theta^{-} \bar{\theta}^{-} (D_0 - D_3)  \Phi_{\tilde{Q}},
\ee
 
The conclusion is that our reduction of the  ${\cal N} = 2, d=4$ theory on a $T^2$ with magnetic flux gives rise to 
a (2,2) supersymmetric theory in 2 dimensions with (2,2) matter chiral multiplets $\Phi_{Q}^{(2,2)}$ and $\Phi_{\tilde{Q}}^{(2,2)}$ .

The coupling between the matter fields and the gauge fields is represented by a (0,2)  superpotential. In \cite{kuta1} a superpotential of
interest for a collection of $\Lambda_a=\phi_- - \sqrt{2} \theta^{+} F$ Fermi superfields and 
$\Phi_i=\phi_i+\sqrt{2} \theta^+ \psi_{+}$ chiral superfields was
\be
\int d^2 x d~\theta^{+} \Lambda_a J^a(\Phi_i) = \int d^2 x (F_a J^a + \psi_{-a} \psi_{+i} \frac{\partial J^a}{\partial \phi_i})
\ee
where $J^a$ are holomorphic functions of the chiral superfields $\Phi_i$. In our case the superpotential inherited from the 
${\cal N} = 2, d=4$ theory couples one Fermi superfield $\Lambda_{\tilde{Q}}$ and two chiral superfields $\Phi_{\Phi}$ and 
$\Phi_{Q}$ of the form:
\be
\int d^2 x d~\theta^{+}  \Lambda_{\tilde{Q}}  \Phi_{\Phi}  \Phi_{Q} = \int d^2 x (F_{\tilde{Q}}  \phi_{\Phi}  \phi_{Q} + 
\psi_{-,\tilde{Q}} \psi_{+,\Phi} \phi_{q} + \psi_{-,\tilde{Q}} \phi_{\Phi} \psi_{+\tilde{Q}})
\ee 
 The term $\psi_{-,\tilde{Q}} \phi_{\Phi} \psi_{+\tilde{Q}}$ provides the usual description of the Coulomb branch related to a vacuum expectation value for the field $\phi$. 

\subsection{From $(2,2), d=2$ to $(0,2), d=2$} 

In four dimensions a theory with  ${\cal N} = 1$ SUSY and fields $Q, \tilde{Q}$ in the (anti) fundamental representations of the gauge group is obtained by adding a general polynomial in  the   ${\cal N} = 1$ chiral multiplet $\Phi$ to the term  $Q~\Phi~\tilde{Q}$:
\be
\sum_{k=1}^{n} \frac{1}{k+1} \mbox{Tr} \Phi^{k+1} + Q~\Phi~\tilde{Q}
\ee
which implies the extremum condition 
\be
\sum_{k=1}^{n} \mbox{Tr} \Phi^{k} + Q~\tilde{Q}= 0.
\ee
This relates the vevs of the flavour fields to the one of the scalar $\Phi$.
 
After compactification to four dimensions and partially breaking the SUSY to (2,2),  in  (0,2) language we can write the potential as
 $\Phi_{\Phi}^n \Lambda_{\Phi}$ so the total superpotential is
\be
\int~d~\theta^+  (\sqrt{2} \Phi_{Q} \Phi_{\Phi}  \Lambda_{\tilde{Q}}+ \Phi_{\Phi}^n \Lambda_{\Phi})
\ee
whose derivative with respect to $\Phi_{\Phi}$ implies
\be
\int~d~\theta^+ (\sqrt{2} \Phi_{Q}   \Lambda_{\tilde{Q}}+ \Phi_{\Phi}^{n-1} \Lambda_{\Phi}) = 0
\ee
The solution of this equation would provide a (0,2) field theory in two dimensions. 

\section{Brane Configurations and Geometries with D and F terms}

\subsection{${\cal N} = 2, d=4, U(N_c)$ theories}

\subsubsection{Geometric Engineering}

We first consider the IIB picture where the gauge group lives on D5 branes wrapped on 2-cycles. The geometry corresponds to a resolved $x y = z^2$ 
singularity where the singular $x=y=z=0$ point is replaced by a  $\mathbb{P}^1$ cycle with normal bundle $O(0) \oplus O(-2)$. By wrapping D5 branes on the $\mathbb{P}^1$ cycle, 
the field theory living on the D5 branes is ${\cal N} = 2, d=4, U(N_c)$. We have a 5 parameter family of deformations in type IIB string theory:

- two choices for a two form composed of  the NS field $B^{NS}$ and the RR field $B^{RR}$. 

- the holomorphic volume of the  $\mathbb{P}^1$ cycle defined by
\be
\label{holo1}
\alpha = \int_{\mathbb{P}^1} \frac{ dx dy}{z}
\ee

- the real Kahler modulus which is the integral of the Kahler form $k$

\be
\label{holo2}
r = \int_{\mathbb{P}^1} k
\ee

The stringy volume of the  $\mathbb{P}^1$ cycle is $V=(B_{NS}^2 + r^2 + \alpha^2)^{1/2}$. 
and the coupling constant of the field theory on the D5 branes is 
\be
\frac{1}{g^2} = \frac{V}{g_s}.
\ee 
The real parameter $r$ is related to the value of D-term (the real value of the 
D field inside the ${\cal N} = 1$ vector multiplet component of the 
${\cal N} = 2$ vector multiplet) and the complex parameter $\alpha$ is related to the 
value of the F-term (the complex value of the F field inside the ${\cal N} = 1$ chiral multiplet $\Phi$ component of the 
${\cal N} = 2$ vector multiplet).

\subsubsection{Brane Configurations}
The T-dual of the geometric picture is a brane configuration containing the following:

- two NS branes oriented along (012345) directions

- $N_c$ D4 branes in the (01236) directions suspended between the NS branes.

The D4 branes can move along the NS branes in the (45) directions, 
spanning the Coulomb branch and preserving the full ${\cal N} = 2$ 
supersymmetry. The coupling constant of the  ${\cal N} = 2, SU(N_c)$ theory is proportional to the distance between the the NS branes in the $x^6$ directions.

What about the directions $x^7, x^8, x^9$? We can move the NS branes with respect to each other in these direction, keeping the D4 branes suspended between the NS branes. The $x^7$ displacement corresponds to the geometric real blow-up 
parameter $r$ of the $D$ and the $x^8, x^9$ displacements to the deformation
complex parameter $\alpha$. The steps from $r=0, \alpha=0$ to $r \ne 0, \alpha \ne 0$ are understood in brane configurations as

a)  $r=0, \alpha=0$ to $r = 0, \alpha \ne 0$ corresponds to separating the NS branes in the  $x^8, x^9$ directions.
 
b)  $r=0, \alpha \ne 0$ to $r \ne 0, \alpha \ne 0$ corresponds to  separating the NS branes in the  $x^7$ directions.

\subsection{SUSY breaking ${\cal N} = 2, d=4$ to ${\cal N} = (2,0), d=2$ for $U(N_c) \times U(N_f)$}

\subsubsection{Geometric Engineering}

We now  consider the $A_2$ singularity $x y = z^3$ and its resolutions. The singular $x=y=z=0$ point is replaced by a two  $\mathbb{P}^1$ cycles with 
overlapping $O(0) \oplus O(-2)$ normal bundles. By wrapping $N_c$ D5 branes on 
the first  $\mathbb{P}^1$ cycle and $N_f$ D5 branes on the second  $\mathbb{P}^1$ cycle, 
the field theory living on their worldvolume is 
${\cal N} = 2, d=4, U(N_c) \times U(N_f)$. Each  $\mathbb{P}^1$ cycle has a 
5 parameter family of deformations in type IIB string theory including the 
NS and RR 2-forms, the holomorphic volumes $\alpha_1, \alpha_2$ and the real 
Kahler parameters $r_1, r_2$:
\be
\label{holo3}
\alpha_i = \int_{ \mathbb{P}^1_i} \frac{ dx dy}{z}, r_i = \int_{\mathbb{P}^1_i} k,~~  i=1,2.
\ee
Besides the gauge multiplets for $U(N_c) \times U(N_f)$ there are also  ${\cal N} = 2$ hypermultiplets which are
collections of ${\cal N} = 1$ chiral multiplets $Q$ and $\tilde{Q}$ transforming in $(N_c, \tilde{N}_f)$ and 
 $(\tilde{N}_c, N_f)$ representation. In this work we consider the  $\alpha_1, \alpha_2$ parameters to be zero, the D term $r_1$ for the gauge group remains at zero whereas the D term $r_2$ for the flavour group is 
non-zero.

In order to partially preserve the supersymmetry we compactify on a torus with magnetic flux.  
When compactifying on a $T^2$, the ${\cal N} = 2, d=4$ theory becomes  ${\cal N} = (4,4), d=2, U(N_c) \times U(N_f)$.  
As we saw in the previous section, turning on a magnetic flux equal to the D term implies that the supersymmetry is 
broken to  ${\cal N} = (2,2), d=2$. Therefore, we consider a fibration of the resolved $A_2$ singularity over a two torus.
 
\subsubsection{SUSY breaking in Brane Configurations}

Consider the compactification two torus to be in the $x^1,x^2$ directions. The above two steps of supersymmetry breaking can 
be described in brane configurations as follows:

1) breaking to $N=(2,2), D=2$.

- having a non-zero D term $r \ne 0$ implies a rotation of the D4 branes in the $x^6, x^7$ plane by an angle $\theta$ such 
$\mbox{tan}~\theta = r$.

- having a non-zero flux $M \ne 0$ implies a rotation of the D4 branes in the $x^1, x^2$  plane by an angle $\theta$ such 
$\mbox{tan}~\theta = M$.
 
The NS branes are left unrotated. If $r = M$, the supersymmetry is partially preserved as (2,2) in 2 dimensions. 

2)  breaking to $N=(2,0), D=2$.

A mass for the adjoint fields $\Phi_1, \Phi_2$ corresponds to rotating the NS branes in the (4589) plane. The $N=(0,2)$ two dimensional configuration is obtained from the  $N=(4,4)$ two dimensional  configuration by rotating the 
D4 branes in the (1267) plane and the NS branes in the (4589) plane. 

\section{Conclusions}
 In this work we covered the steps describing the breaking of   ${\cal N} = 2$ supersymmetry in four dimensions to (0,2) 
supersymmetry in two dimensions.
The theories live on D5 branes wrapped on 2-cycles inside Calabi-Yau 3-folds or 4-folds and the supersymmetry is partially broken after a further compactification on 
a two torus with magnetic flux. The magnetic flux is made equal to the volume of the 2 cycle to preserve (2,2) in two dimensions and a further 
deformation of the normal bundle to the 2-cycle leads to (0,2) theory in two dimensions. We left a collection of issues for future publications. On one hand the 
line bundles considered in this work are of zero degree and it is important to generalise to bundles of even degree (we need an even number to allow the 
definition of the square root bundle). It is also important to understand the different types of geometric deformations as line bundles over the two torus.

\section*{Acknowledgments}

This work was supported in part by STFC. We would like to thank Atif Choudri and Nicola Pagani for discussions.

\end{document}